\documentclass[twocolumn]{aastex701}

\usepackage{gensymb} % for degree symbol

\newcommand{\msun}{\,M$_\odot$}
\newcommand{\hi}{\ion{H}{1}}
\newcommand{\kms}{\,km\,s$^{-1}$}

\begin{document}

\title{The First RELHIC? Cloud-9 is a Starless Gas Cloud\footnote{Based on observations made with the NASA/ESA Hubble Space Telescope, obtained at the Space Telescope Science Institute, which is operated by the Association of Universities for Research in Astronomy, Inc., under NASA contract NAS5-26555. These observations are associated with program 17712.}} 

\author[0000-0002-5259-2314]{Gagandeep S. Anand}
\affiliation{Space Telescope Science Institute, 3700 San Martin Drive, Baltimore, MD 21218, USA}
\email{ganand@stsci.edu}  

\author[0000-0001-8261-2796]{Alejandro Benítez-Llambay}
\affiliation{Dipartimento di Fisica G. Occhialini, Università degli Studi di Milano Bicocca, Piazza della Scienza, 3 I-20126 Milano MI, Italy}
\email{alejandro.benitezllambay@unimib.it}

\author[0000-0002-1691-8217]{Rachael Beaton}
\affiliation{Space Telescope Science Institute, 3700 San Martin Drive, Baltimore, MD 21218, USA}
\email{rbeaton@stsci.edu}  

\author[0000-0003-0724-4115]{Andrew J. Fox}
\affiliation{AURA for ESA, Space Telescope Science Institute, 3700 San Martin Drive, Baltimore, MD 21218, USA}
\email{afox@stsci.edu}  

\author[0000-0003-3862-5076]{Julio F. Navarro}
\affiliation{Department of Physics and Astronomy, University of Victoria, Victoria, BC V8P 5C2, Canada}
\email{jfn@uvic.ca}

\author[0000-0003-2676-8344]{Elena D'Onghia}
\affiliation{Department of Astronomy, University of Wisconsin, Madison, WI 53706, USA}
\email{edonghia@astro.wisc.edu}

\begin{abstract}
Five-hundred-meter Aperture Spherical Telescope (FAST) observations have recently identified a compact
\hi\ cloud (hereafter Cloud-9) in the vicinity of the spiral galaxy M94. This identification has been confirmed independently by Very Large Array (VLA) and Green Bank Telescope (GBT) observations.
Cloud-9 has the same recession velocity as M94, and is therefore at a similar distance ($\sim$4.4~Mpc). It is compact ($\sim1\arcmin$ radius, or $\sim 1.4$ kpc), dynamically cold ($W_{50}=12$\kms), 
non-rotating, and fairly massive, with an \hi\ mass of $\sim 10^6$\msun.
Here we present deep Hubble Space Telescope/Advanced Camera for Surveys (HST/ACS) imaging designed to search for a luminous stellar counterpart. We visually rule out the presence of any dwarf galaxy with stellar mass exceeding 10$^{3.5}$\msun. A more robust color-magnitude diagram-based analysis conservatively rules out a 10$^{4}$\msun\ stellar counterpart with $99.5^{+0.5}_{-8.2}$$\%$ confidence.\ The non-detection of a luminous component reinforces the interpretation that this system is a Reionization-Limited \hi\ Cloud (RELHIC); i.e.,  
a starless dark matter halo filled with hydrostatic gas in 
thermal equilibrium with the cosmic ultraviolet background.
Our results make Cloud-9 the leading RELHIC candidate of any known compact \hi\ cloud.
This provides strong support for a cornerstone prediction of the $\Lambda$CDM model, 
namely the existence of gas-filled starless dark matter halos on sub-galactic 
mass scales, and constrains the present-day threshold halo mass for galaxy formation.
\end{abstract}

%%%%%%%%%%%%%%%%%%%%%%%%%%%%%%%%%%%%%%%%%%%%%%%%%%%%%%%%%%%%%%%%%%
\section{Introduction}\label{sec:intro}
%%%%%%%%%%%%%%%%%%%%%%%%%%%%%%%%%%%%%%%%%%%%%%%%%%%%%%%%%%%%%%%%%%

 A fundamental prediction of the Lambda Cold Dark Matter ($\Lambda$CDM) model is the formation of dark matter halos over a vast range of scales, with a mass function that rises steeply towards low masses~\citep{Press1974, Jenkins2001}. The abundance of halos far exceeds that of known galaxies, implying that not all halos are able to host luminous galaxies. This has been interpreted to mean that galaxies only form in halos that exceed a ``critical" mass, $M_{\rm crit}(z)$, a result supported by numerical simulations and theoretical modeling
~\citep[e.g.,][and references therein]{Blumenthal1984,Hoeft2006, Okamoto2009, Sawala2016,  Benitez-Llambay2020}. After reionization, this threshold is set by the balance between gravity, gas cooling, and photoheating by the cosmic ultraviolet background (UVB). At the present epoch, $M_{\rm crit} \approx 10^{9.7} \ M_{\odot}$~\citep[e.g.][]{Benitez-Llambay2020, Nebrin2023}.

Halos below $M_{\rm crit}$ do not retain much photoheated gas and, unless they were able to form stars before reionization, remain starless. By contrast, halos above $M_{\rm crit}$ are massive enough to overcome thermal pressure, enabling galaxy formation. Halos just below $M_{\rm crit}$ occupy an intermediate regime: they are not massive enough to form stars today, but massive enough to retain some of their gas. Of these, the most mass massive retain enough baryons in their central regions for hydrogen recombination to occur. These massive, \ion{H}{1}-rich, starless systems were termed Reionization-Limited \ion{H}{1} Clouds (RELHICs) by~\cite{Benitez-Llambay2017}.

\begin{figure} 
\begin{center}
\includegraphics[width=0.4\textwidth]{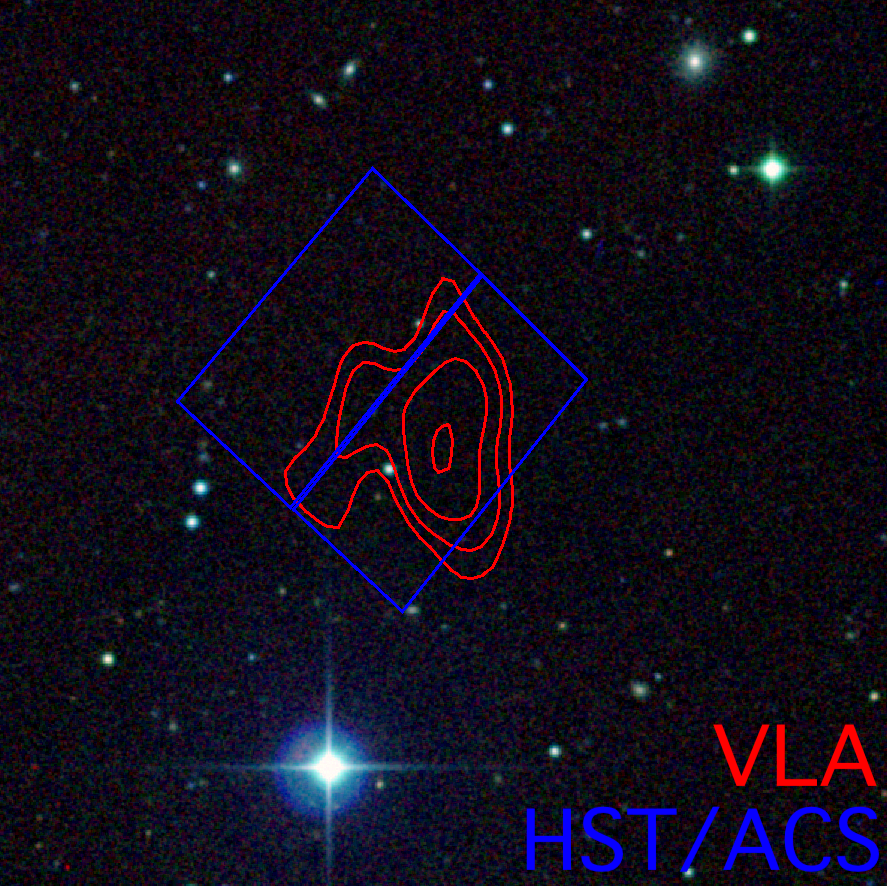}
\end{center}
\caption{Digitized Sky Survey image covering a 10$'\times$10$'$ region around Cloud-9. The VLA \hi\ contours \citep{Benitez-Llambay2024} are shown in red, and the footprint of our HST/ACS observations are shown in blue.}
\label{fg:footprint} 
\end{figure} 

Cosmological hydrodynamical simulations suggest that, despite lacking stars, RELHICs should host compact, nearly spherical cores of neutral hydrogen \citep{2016MNRAS.457..844F, 2016MNRAS.457.1931S, Benitez-Llambay2017}. The gas is in hydrostatic equilibrium with the dark matter and in thermal equilibrium with the UVB. These properties make RELHICs detectable through their narrow 21 cm emission line, with characteristic velocity widths of $W_{50} \lesssim 20 \ \rm km \ s^{-1}$. These simulations also indicate that RELHICs are susceptible to ram-pressure stripping, and, therefore, are unlikely to be found very close to a massive galaxy.

Because their gas is pressure-supported, RELHICs offer a unique window into the underlying dark matter distribution, bypassing the complexities of galaxy formation that often prevent robust inferences on the nature of dark matter~\cite[see e.g.,][for a review]{Bullock2017}.
Detecting RELHICs would be a breakthrough, providing clear evidence for bound, collapsed dark matter halos on sub-galactic scales, and offering constraints on the nature of dark matter in a previously unexplored regime.

A few RELHIC candidates have recently been identified with the Five-hundred-meter Aperture Spherical Telescope (FAST), which is well-suited for surveying large areas of the northern sky in \hi . An example is FAST J0139+4328~\citep{Xu2023}, which has no detected luminous counterpart yet contains an \ion{H}{1} mass of $\approx 8.3 \times 10^{7} \ M_{\odot}$. However, the double-peaked shape of its \hi\ profile indicates that the gas is rotationally supported, suggesting that the gas has collapsed into a disk, and that J0139+4328 may actually contain a faint galaxy. 
Indeed, the current upper limit on the stellar mass of any luminous counterpart is only $\approx 7 \times 10^5 \ M_{\odot}$. This limit should be improved by deeper imaging before it can be claimed that J0139+4328 is actually ``dark".

Other notable candidates include some of the {\tt ALFALFA} Ultra Compact High-Velocity Clouds cataloged by~\cite{Adams2013}. However, most of the candidate clouds are likely associated with the Milky Way disk, and, therefore, not RELHICs, which could not survive the ram-pressure stripping due to Galactic gas. In addition, those clouds not associated with the Milky Way are either too extended, too irregular, or exhibit line widths too broad to be consistent with RELHICs ~\citep{Benitez-Llambay2017}. The scarcity of robust candidates is perhaps unsurprising: the relatively shallow depth of {\tt ALFALFA} limits sensitivity to the low fluxes predicted for RELHICs, while the large beam of the Arecibo radio telescope prevents them from being spatially resolved.

Among all candidate detections to date, one stands out as a particularly compelling: Cloud-9, recently identified in the vicinity of M94 with FAST by~\cite{Zhou2023}. Unlike previous dark \ion{H}{1} cloud candidates, Cloud-9 exhibits properties consistent with RELHICs: its recessional velocity (304 \kms) is positive and similar to that of M94; its 21 cm line profile has a width of $W_{50} \lesssim 20 \ \rm km \ s^{-1}$, and its total \ion{H}{1} flux is consistent with that expected for a RELHIC at the M94 distance~\citep{Zhou2023, Benitez-llambay2023,  Karunakaran2024, Benitez-Llambay2024}. The small angular separation from M94 ($\sim$ 70 kpc in projection) and comparable recessional velocity suggest a physical association, providing a lower limit on their relative distance. 

Higher-resolution VLA observations (see Figure \ref{fg:footprint}) further support the idea that 
Cloud-9 is at M94's distance: the column density isocontours––which appeared round in FAST data-- appear slightly distorted in the VLA image, possibly indicating ram-pressure interactions with the gaseous halo of M94~\citep{Benitez-Llambay2024}.

Detailed analysis of Cloud-9 indicates that, at the assumed distance of M94, the system requires a substantial amount of gravitational mass beyond its \ion{H}{1} content to remain in hydrostatic equilibrium~\citep{Benitez-llambay2023, Benitez-Llambay2024}. Interpreting this additional mass as cold dark matter yields a tight constraint on the total halo mass of $\sim$$5\times 10^{9} \ M_{\odot}$, remarkably close to $M_{\rm crit}$ at the present day. Moreover, a comparison with mock RELHIC models shows that Cloud-9's properties are broadly consistent with a $\Lambda$CDM RELHIC, with a total neutral hydrogen mass of $M_{\rm HI} \sim 10^6 \ M_{\odot}$~\citep{Benitez-llambay2023}. Together, these results make Cloud-9 the firmest RELHIC candidate known to date.

A key challenge to interpret Cloud-9 as a RELHIC lies in the loose upper limit on its stellar content available at present. Its \ion{H}{1} mass and radial distribution closely resemble those of Leo T, a  dwarf galaxy in the outskirts of the Milky Way~\citep[][]{Irwin2007, Benitez-Llambay2024}. Notably, Leo T's stellar mass ($\sim$$10^{5} \ M_{\odot}$) is comparable to the current upper limit for a luminous counterpart to Cloud-9, based on imaging from the DESI Legacy Survey~\citep{Zhou2023}. This makes it difficult to exclude the presence of a Leo T-like stellar counterpart. Confirming the RELHIC nature of Cloud-9 therefore requires deeper observations capable of significantly lowering this upper limit.

In this Letter, we present deep HST/ACS imaging of Cloud-9, obtained to search explicitly for a stellar counterpart.
In Section~\ref{sec:obs} we describe the HST observations and data reduction.
In Section~\ref{sec:cmd} we place limits on the stellar mass of a stellar counterpart to Cloud-9. We discuss our results in Section~\ref{sec:discussion} and use Section ~\ref{sec:summary} to provide a brief summary.

%%%%%%%%%%%%%%%%%%%%%%%%%%%%%%%%%%%%%%%%%%%%%%%%%%%%%%%%%%%%%%%%%%
\section{HST Observations and Reductions}\label{sec:obs}

\begin{figure*} 
\begin{center}
\includegraphics[width=0.75\textwidth]{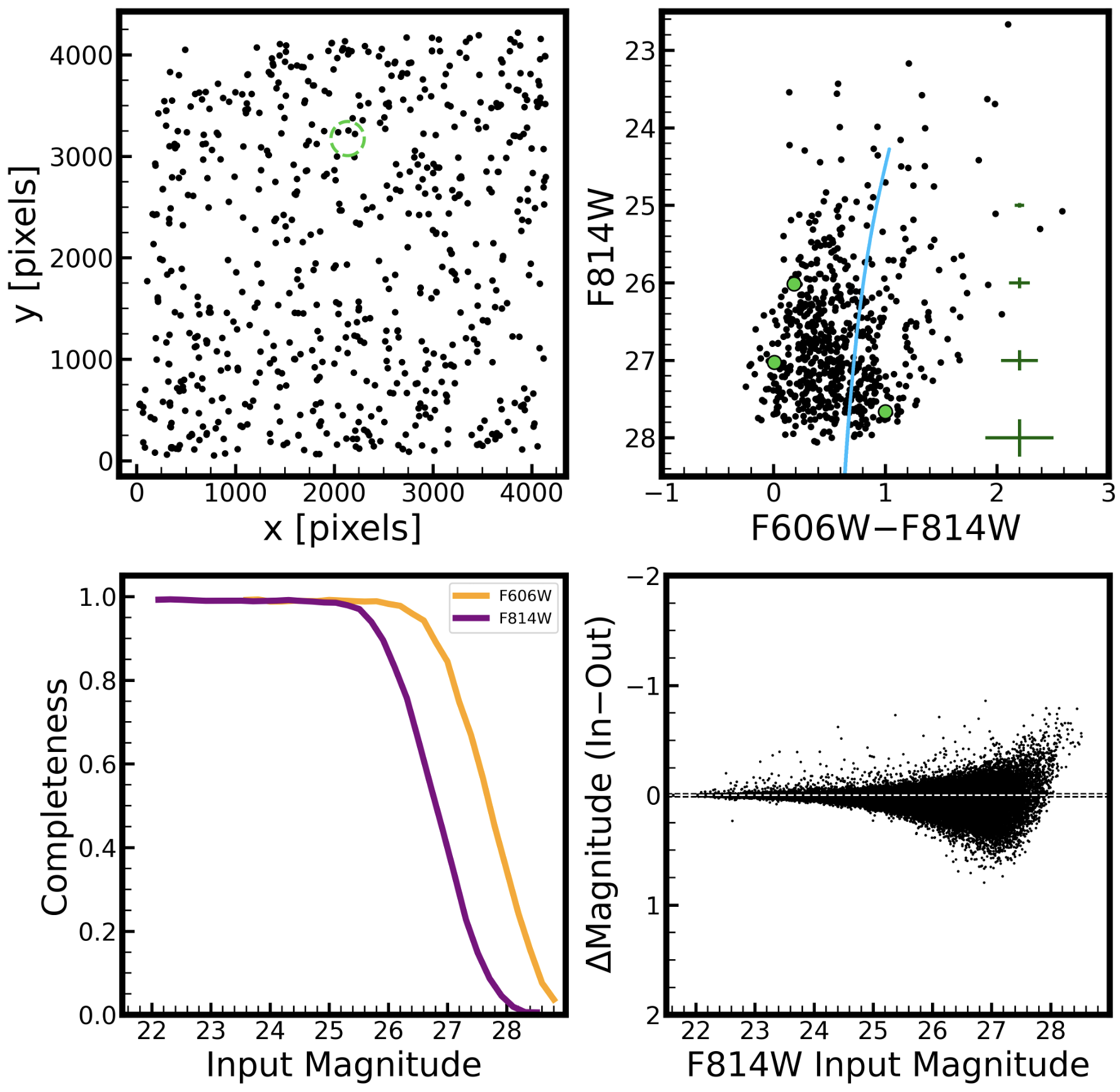}
\end{center}
\caption{\textbf{Top Left:} The spatial distribution of sources fulfilling our quality criteria. The green dashed circle indicates the effective radius (8.4$''$ or 180~pc projected) of a Leo~T analog at the distance of M94, and is centered at the location of Cloud-9. \textbf{Top Right:} CMD of the field. The three sources within the green circle in the top panel are shown in green. We also show an RGB isochrone for an old, metal-poor stellar population (10~Gyr, [Fe/H] = $-$2.0~dex) drawn from MIST, and placed at the distance of M94. The green crosses on the right-hand side indicate typical photometric uncertainties for various F814W magnitudes at the color of the RGB isochrone. \textbf{Bottom Left:} Photometric completeness distributions for both of our filters, as determined by the artificial star experiments. \textbf{Bottom Right:} Difference between input and output magnitudes from the artificial star experiments for the primary CMD filter (F814W).}
\label{fg:xy+cmd} 
\end{figure*}

\begin{figure*} 
\begin{center}
\includegraphics[width=\textwidth]{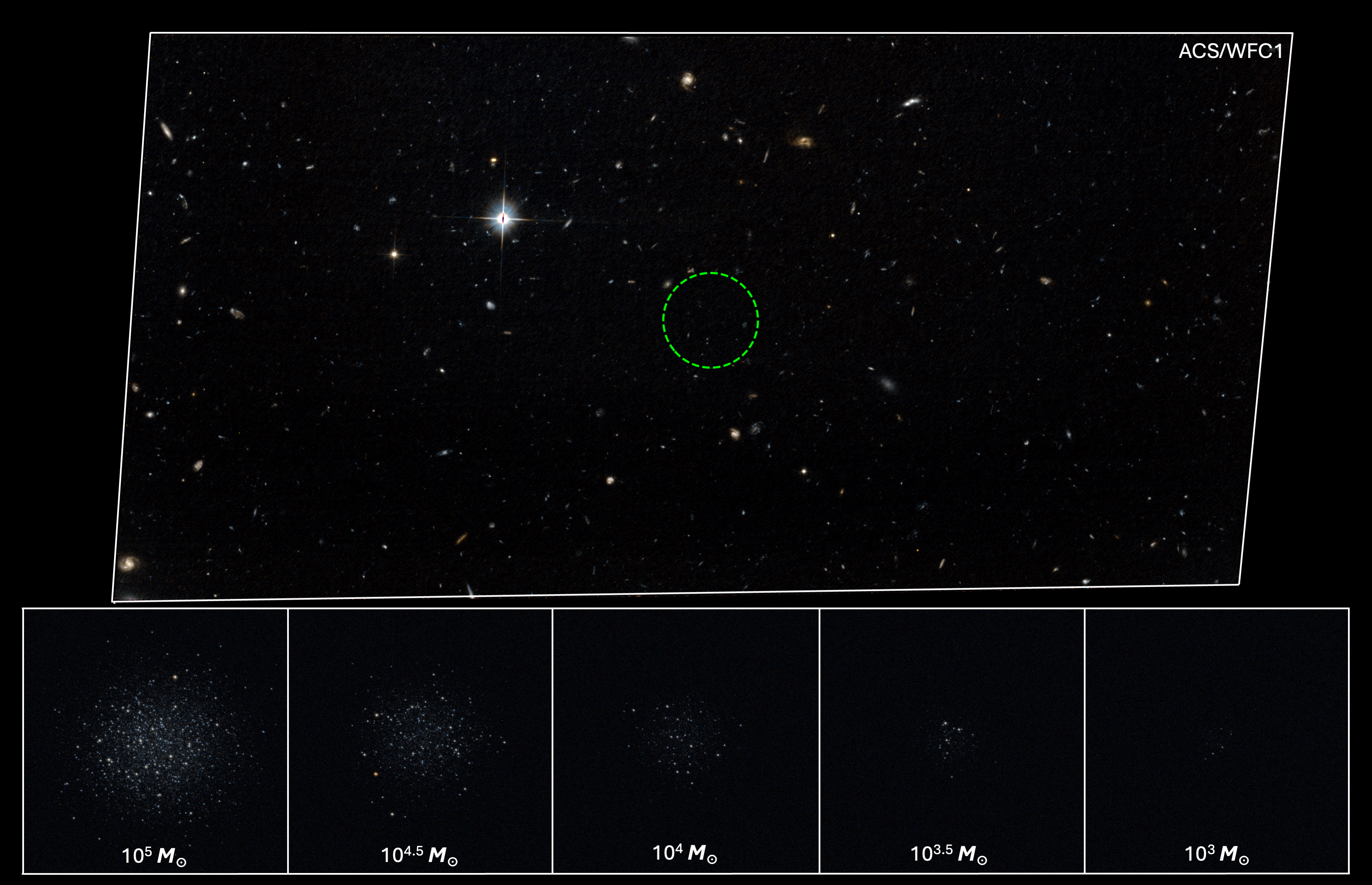}
\end{center}
\caption{\textbf{Top:} A color composite of our ACS/WFC1 imaging ($\sim$202$''$$\times$100$''$). The green circle ($r=8.4$$''$) marks the VLA \hi\ maximum column density, with a radius corresponding to the effective radius of a Leo~T analog at the distance of M94. \textbf{Bottom:} Simulated dwarf galaxies spanning a range of stellar masses, generated under the same observing conditions and shown at the same physical scale as the ACS data. A Leo~T analog ($M_{\star}=10^{5} \ M_{\odot}$) would be readily detected, but no stellar counterpart is visible down to at least $M_{\star}\sim10^{3.5}M_{\odot}$.}
\label{fg:artpop_galaxy_images} 
\end{figure*} 

We obtained observations with the Hubble Space Telescope (HST) Advanced Camera for Surveys Wide-Field Channel (ACS/WFC) as part of program GO-17712 in Cycle 32~\citep{2024HSTProposal}. The observations were taken in four visits of two orbits each between 17 and 19 February 2025, and were designed to target a depth of four magnitudes below the tip of the red giant branch (TRGB) at the distance of M94 ($d_{\rm M94}=4.41 \rm \ Mpc$; \citealt{2021AJ....162...80A}).

The nominal position of Cloud-9 (based on the VLA \hi\ maximum) was placed at the center of the WFC1 chip (see Figure \ref{fg:footprint}). Each visit employed a small, 5-pixel dither to allow rejection of cosmic rays and other detector artifacts. A loss of guide stars occurred near the end of the final F814W exposure during the first visit, truncating that frame, but otherwise the dataset was unaffected. In total, the observations comprise 9036 s in the F606W filter (``wide-V"), and 8749.4 s in the F814W filter (approximately $I$-band). 

Because our data were obtained in four separate visits, we first aligned all underlying exposures using the charge-transfer corrected \texttt{FLC} files. For the F606W images, we applied the \texttt{tweakreg} procedure from the {\tt DrizzlePac} software suite \citep{2015ASPC..495..281A} to align the individual exposures, and then combined them into a deep ``drizzled'' image with {\tt AstroDrizzle} (also within the DrizzlePac package). We repeated this process independently for the F814W data. We then mutually aligned the two drizzled images, and used \texttt{tweakback} to back-propagate this improved alignment to the original \texttt{FLC} files. Using these adjusted \texttt{FLC} exposures, we again created drizzled images of the fields in the separate filters, ensuring consistent alignment across the full dataset. 

We performed PSF photometry with the {\tt DOLPHOT} stellar photometry package \citep{2000PASP..112.1397D, Dolphin2016}, adopting the reduction parameters recommended by the PHAT program \citep{2014ApJS..215....9W,2023ApJS..268...48W} for ACS/WFC imaging. These parameters have been extensively validated across a wide range of extragalactic fields. The F814W drizzled image served as the undistorted reference frame, while the PSF photometry was carried out on the individual \texttt{FLC} images to avoid issues introduced by resampling of the PSF in the drizzled images. All magnitudes are reported in the Vega photometric system.

The base photometric output was culled using quality cuts––we tested several iterations from the literature~\citep{2014ApJS..215....9W, 2021AJ....162...80A, 2023RNAAS...7...23W,2024ApJS..275....5M}, and found that the following relatively strict cuts provided the most effective rejection of artifacts, such as spurious sources found in the diffraction spikes of bright foreground stars: \texttt{Crowd} $<$ 0.5 (both filters), \texttt{|Sharp|} $<$ 0.1 (both filters), \texttt{Type} $\le$ 1, \texttt{S/N} $\ge$ 5 (both filters), and \texttt{Flag} $\le$ 2 (both filters). The final catalog contains 667 sources.

To assess photometric bias, completeness, and measurement uncertainties, we conducted artificial star experiments with {\tt DOLPHOT}. We injected and recovered approximately 500,000 artificial stars one at a time using the same reduction setup. The same quality cuts were applied to the resulting artificial star catalog. The bottom panels of Figure \ref{fg:xy+cmd} show our photometric completeness distributions in both filters, as well as the differences between input and output magnitudes for artificial stars in our primary filter (F814W).

The color-magnitude diagram (CMD) of the full ACS field, together with the spatial distribution of detected sources, is shown in Figure \ref{fg:xy+cmd}. This CMD closely resembles those derived from other ``background" fields, such as the Hubble eXtreme Deep Field \citep{2017ApJ...836...74J}. In the top panel, a green dashed circle marks the location of the VLA \hi\ maximum, with a radius of 8.4'', corresponding to the effective radius of Leo T ($180$ pc) at the distance of M94. In the bottom panel, green symbols highlight the (only) three sources that fall within this region, alongside a 10 Gyr, metal-poor ([Fe/H] = $-$2.0) red giant branch (RGB) isochrone at this distance, drawn from MIST \citep{2016ApJS..222....8D,2016ApJ...823..102C}. The green crosses on the right indicate the observed photometric uncertainties for stars spanning a range of magnitudes at the color of the isochrone.

%%%%%%%%%%%%%%%%%%%%%%%%%%%%%%%%%%%%%%%%%%%%%%%%%%%%%%%%%%%%%%%%%%

%%%%%%%%%%%%%%%%%%%%%%%%%%%%%%%%%%%%%%%%%%%%%%%%%%%%%%%%%%%%%%%%%%
\section{Searching for a Stellar Counterpart}\label{sec:cmd}
%%%%%%%%%%%%%%%%%%%%%%%%%%%%%%%%%%%%%%%%%%%%%%%%%%%%%%%%%%%%%%%%%%

\subsection{Visually Ruling Out a Leo T analog}

With our deep ACS imaging and photometry in hand, we examined the field for a stellar counterpart to Cloud-9. Using the {\tt ArtPop} software package \citep{2022ApJ...941...26G}, we generated simulated images of dwarf galaxies at the distance of M94 under our ACS/WFC observing conditions, including the telescope parameters, filter set, exposure times, photometric zeropoints, and point-spread functions. 

Following the stellar mass-size relation for nearby dwarf satellites of Milky-Way analogs (including M94) from \cite{2021ApJ...922..267C}, we adopted as a starting point a Leo T analog with $M_{\star} =  10^{5} \ M_{\odot}$, and size, $r_{\rm eff} = 180 \rm \ pc$ \citep{2007ApJ...656L..13I,2012ApJ...748...88W}, and decreased the stellar mass by factors of $\sim 3$ down to $M_{\star} = 10^{3} \ M_{\odot}$ (with $r_{\rm eff} = 57 \rm \ pc$). We assumed an old, metal-poor stellar population of age 10 Gyr and [Fe/H] = $-$2.0, noting that the precise metallicity has little impact on the results. We also note that the adoption of an old stellar population is the most conservative, as a stellar counterpart with any fraction of its stars in a young or intermediate age would produce even more numerous and bright stars that would be visible in our HST observations.

The top panel of Figure~\ref{fg:artpop_galaxy_images} shows our ACS/WFC1 imaging, centered on the VLA \hi\ centroid. The bottom panel shows the artificially observed dwarf galaxies, shown at the same physical scale as the ACS data. It is clear that a Leo T analog with $M_{\star}\sim 10^{5} \ M_{\odot}$ would be readily visible if present. Such a counterpart would produce dozens of resolved stars above our magnitude limit (see the HST color-magnitude diagram in \citealt{2012ApJ...748...88W}). In contrast, no comparable resolved structure is present in the image, with non-detections extending down to $M_{\star}\sim 10^{3.5} \ M_{\odot}$ or lower (though we caution that stochasticity in the number of bright stars is significant at these low stellar masses). While these simulations do not capture all the intricate detector effects associated with ACS---such as charge-transfer efficiency losses or bias striping---these effects are largely corrected by the ACS pipeline. Importantly, the simulations are intended primarily as a visual tool, illustrating what a hypothetical stellar counterpart to Cloud-9 would look like if present in our data. In the next subsection, we introduce a more quantitative framework for our analysis.

\subsection{Stellar Population Simulations}

Given the absence of a visible stellar counterpart to Cloud-9 in our HST imaging, we next quantify the limits on its underlying stellar mass. A system with $M_{\star}=10^{5} \ M_{\odot}$ would be unmistakable, but we also considered a possible $M_{\star}=10^{4} \ M_{\odot}$ counterpart. While many star-formation history codes (e.g. \citealt{2002MNRAS.332...91D, 2011AJ....141..106J, 2023ApJ...943..139M}) can fit for a total stellar mass in their procedure, they are not designed to determine upper limits for stellar mass content in the apparent absence of stars. Instead, we used {\tt ArtPop} \citep{2022ApJ...941...26G} to simulate stellar (Vega) magnitudes for a given input stellar mass in our ACS filters, incorporating the observational uncertainties and completeness measured from our artificial star experiments. We adopted the same stellar population parameters (10~Gyr old population with Fe/H = $-$2.0 at the distance of M94) and generated 10,000 CMD realizations to capture stochastic variations at this relatively low stellar mass. To mimic the conditions of our HST observations, we applied our full observational selection functions––photometric bias, completeness, and uncertainties–– to each simulated CMD. 

Turning to the observed CMD, we detect three nominal sources within the 8.4$''$ region centered on Cloud-9. To account for the $\sim 9$'' uncertainty in the position of the \hi\ centroid, we shift the center of our nominal region within this positional uncertainty $\sim200,000$ times, yielding a mean of 3.5$\pm$1 sources associated with Cloud-9. However, because of the large population of unresolved background galaxies (or substructures within them), these sources may not be genuine stars. To quantify the background source population, we use the opposing (WFC2) chip. We place $\sim$200,000 same-sized regions across the chip, shifting their centers to cover the full WFC2 area (avoiding chip edges). This yields a mean of 3.7$\pm$2 background sources per region. Since we observe 3.5$\pm$1 sources in the Cloud-9 region, this corresponds to an overdensity of $-$0.2$\pm$2.2 sources associated with Cloud-9, consistent with no excess at the location of the VLA \hi\ maximum.

\begin{figure} 
\begin{center}
\includegraphics[width=0.45\textwidth]{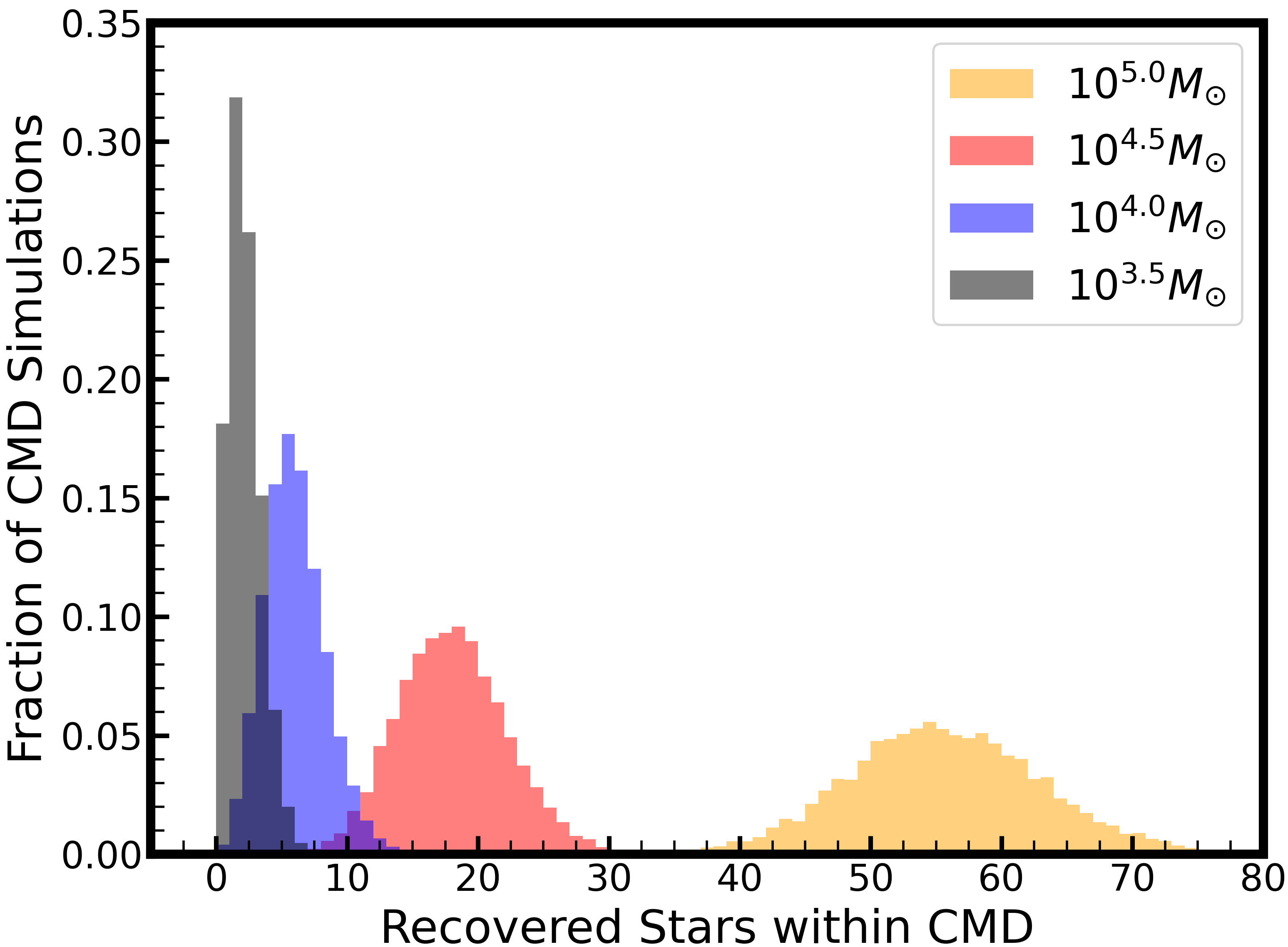}
\end{center}
\caption{Histograms of 10,000 simulations performed for each of a range of input stellar masses (shown in the legend). The x-axis values show the remaining number of visible stars in the simulated CMDs after applying our observational systematics.}
\label{fg:histogram} 
\end{figure}

Figure \ref{fg:histogram} shows the number of stars recovered in 10,000 simulation iterations, performed separately for a range of input stellar masses. For a population with $M_{\star}=10^{4} \ M_{\odot}$, a source was recovered in $99.5\%$ of cases, allowing us to rule out such a population with that confidence level. To incorporate the uncertainty in the derived overdensity ($-$0.2$\pm$2.2 sources) we proceed as follows. On the high end, the maximum overdensity corresponds to 2.0 sources; in the same simulations, 8.7$\%$ of realizations yield two or fewer stars. On the low end, a source population of $-$2.4 sources is nonphysical, and thus excluded with 100$\%$ confidence. Taken together, this results in a $99.5^{+0.5}_{-8.2}$$\%$ confidence limit against a $M_{\star}=10^{4} \ M_{\odot}$ stellar counterpart. Finally, we stress that these limits remain conservative: a genuine stellar counterpart would produce additional resolved or semi-resolved stars that fail our CMD-selection criteria. No such population is observed, further strengthening the case against a stellar counterpart to Cloud-9.

We can also frame the question in terms of maximum stellar mass consistent with our observations. By running simulations across a wide range of stellar masses, we find that a population with $M_{\star}=10^{3.49}\ M_{\odot}$ best matches the data at the nominal center: after including observational biases, incompleteness, and photometric errors, such a system yields on average one or fewer RGB stars, consistent with our observations. We therefore adopt $M_{\star}=10^{3.5}\ M_{\odot}$ as the baseline upper limit for any stellar counterpart to Cloud-9.

Finally, we note that ram-pressure interactions can in principle induce a spatial offset between the stellar and gaseous components of low-mass galaxies. However, the \hi\ morphology of Cloud-9 is relatively regular with only mild asymmetry, suggesting that any such effect would be small. To test this possibility directly, we searched for clusters of point sources within a radius of 1~kpc (47$''$) from the VLA \hi\ maximum, comparable to the extent of the compact~\hi~core (for reference, the top-to-bottom angular size of the WFC1 chip shown in Figure \ref{fg:artpop_galaxy_images} is $\sim$100$''$). In this region we identified a cluster of nine sources. However, given that the search covers a large fraction of the WFC1 chip, such local maxima are expected from the underlying background galaxy clustering. A control search on the opposing WFC2 chip from its center outwards yields two non-overlapping clusters of nine sources, confirming that the WFC1 cluster is not a statistically significant overdensity.

%Finally, we consider the possibility that a stellar counterpart may be displaced from the VLA center. The \hi\ isocontours, while mildly asymmetric, show no signs of strong perturbations, making a large offset between gas and stars unlikely.
%Still, to test this possibility, we searched for clusters of point sources within a radius of 1~kpc (47$''$) from the VLA \hi\ maximum, comparable to the extent of the compact~\hi~core (for reference, the top-to-bottom angular size of the WFC1 chip shown in Figure \ref{fg:artpop_galaxy_images} is $\sim$100$''$). In this region we identified a cluster of nine sources. However, given that the search covers a large fraction of the WFC1 chip, such local maxima are expected from the underlying background galaxy clustering. A control search on the opposing WFC2 chip from its center outwards yields two non-overlapping clusters of nine sources, confirming that the WFC1 cluster is not a statistically significant overdensity. 

%%%%%%%%%%%%%%%%%%%%%%%%%%%%%%%%%%%%%%%%%%%%%%%%%%%%%%%%%%%%%%%%%%
\section{Discussion}\label{sec:discussion}
%%%%%%%%%%%%%%%%%%%%%%%%%%%%%%%%%%%%%%%%%%%%%%%%%%%%%%%%%%%%%%%%%%

\begin{figure*} 
\begin{center}
\includegraphics[]{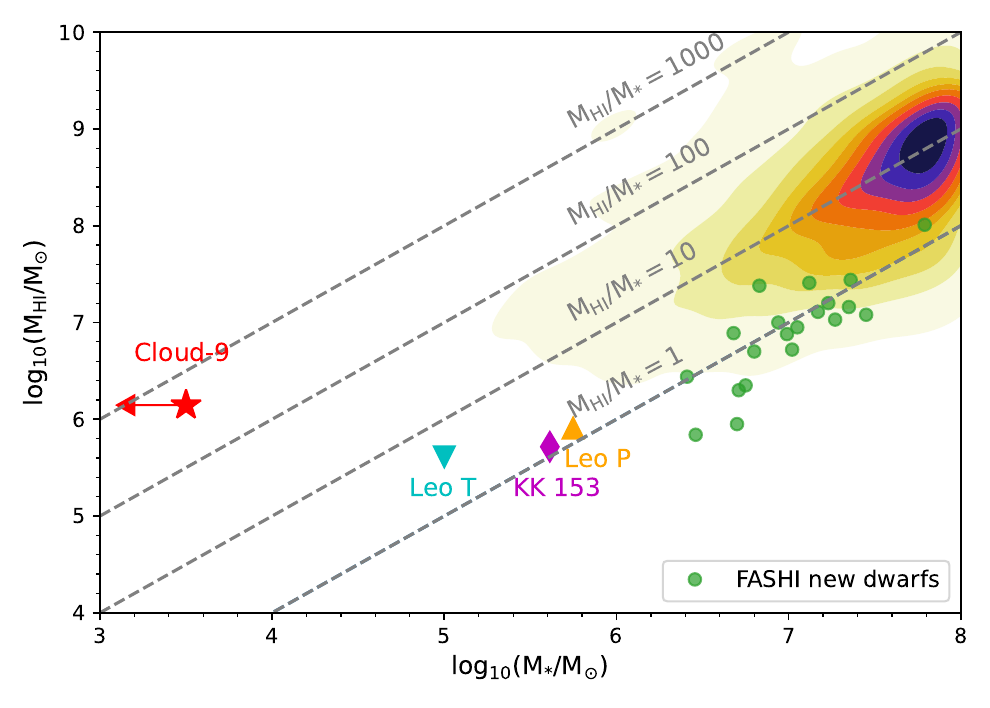}
\end{center}
\caption{\hi\ -stellar mass relation for the ALFALFA-SDSS catalog~\citep[isocontours;][]{Durbala2020}. Newly identified dwarf galaxies in the local volume detected with FAST are shown as green points~\citep{Karachentsev2024}. For comparison, Leo T, KK 153, Leo P, and Cloud-9 are also shown. The stellar mass upper limit for Cloud-9 (red star symbol) underscores its extreme gas richness relative to its possible stellar mass, if interpreted as an ultra-faint galaxy.}
\label{fg:hi-to-stellarmass} 
\end{figure*}

The absence of a faint stellar counterpart associated with Cloud-9 is striking and reinforces its identification as a likely RELHIC. While isolated ultra-faint dwarfs exist with stellar masses comparable to, or below, the upper limit from our analysis, none host such a substantial neutral hydrogen reservoir. Adopting a total \hi\ mass of $M_{\rm HI} \approx 1.4 \times 10^{6} \ M_{\odot}$ from GBT observations~\citep{Karunakaran2024}, together with our stellar mass upper limit, implies an \hi-to-stellar mass ratio of $M_{\rm HI}/M_{\star} \gtrsim 443$, exceeding by orders of magnitude the ratios typical of dwarf galaxies. For comparison, Leo T, KK 153~\citep{Xu2025}, and Leo P~\citep{McQuinn2015} have similar \hi\ masses but stellar counterparts yielding $M_{\rm HI}/M_{\star} \lesssim  4$, consistent with typical values for gas-rich dwarfs~\cite[see, e.g.][]{Lelli2022, Karachentsev2024}. Figure~\ref{fg:hi-to-stellarmass} places Cloud-9 in this context, showing it relative to Leo T and KK 153, newly detected dwarfs with FAST~\citep{Karachentsev2024}, and the ALFALFA-SDSS catalog~\citep{Durbala2020}.

One can consider alternative interpretations for Cloud-9. One possibility is that it is a foreground high-velocity cloud (HVC) of the Milky Way that happens to be projected close on the sky to M94. This scenario, is disfavored by Cloud-9's recessional velocity, which closely matches M94's velocity, as well as the absence of foreground HVCs with comparable velocities in this region of the sky. 
The closest high-velocity clouds to M94 are Complexes M, C, and K \citep{wakker1991, Westmeier2018}, but they have all negative velocities ($v_{\rm LSR}\approx-$200 to $-$100\kms), whereas M94 and Cloud-9 are located at $v_{\rm LSR}\approx+$300\kms.
A second (related) possibility is that Cloud-9 could be associated with the Magellanic Stream, at a distance of $\approx$50--100 kpc. Such an explanation has been proposed for other compact \hi\ HVCs observed at high latitude \citep{2006MNRAS.365.1277V}. However, the Stream lies on the opposite side of the sky as M94, with its closest point (the Leading Arm) around 70 degrees away, rendering this explanation unlikely for Cloud-9.
%Note: 70 degrees is the angular separation between longitude,latitude points 
%270,30 (Leading Arm) and 123, 76 (Cloud-9)
A third possibility is that Cloud-9 is a neutral gas cloud in pressure equilibrium with M94's hot circumgalactic medium. For a cloud at $\approx 10^4 \rm \ K$, confinement by ambient gas at $\sim 10^6 \ K$ requires a density contrast of $\sim 100$. Cloud-9's mean density, $\sim 2\times 10^{-2} \rm \ cm^{-3}$, would then imply a circumgalactic density of $\sim 2\times 10^{-4} \rm \ cm^{-3}$, which is a plausible value. Thus, static pressure confinement cannot be excluded on simple pressure-balance grounds. However, such an equilibrium configuration is unlikely to be long-lived: Cloud-9's perturbed \hi\ morphology suggests ongoing ram-pressure interactions, and unless we are observing it at a very particular moment, a purely non-self gravitating, pressure-confined cloud of this size would be disrupted within a few tens of Myrs~\citep[e.g.][]{Klein1994}.

These arguments lead us to favor the RELHIC or ultra-faint dwarf interpretation over these alternatives. The non-detection of a stellar counterpart, the extreme \hi-to-stellar mass ratio, and the consistency with the predicted mass threshold 
strongly support the RELHIC interpretation.  
Therefore, we conclude that our HST observations make Cloud-9 the most compelling RELHIC candidate to date, and a rare case highlighting the physics that govern the onset of galaxy formation.

%%%%%%%%%%%%%%%%%%%%%%%%%%%%%%%%%%%%%%%%%%%%%%%%%%%%%%%%%%%%%%%%%%
\section{Summary and Future Work}\label{sec:summary}
%%%%%%%%%%%%%%%%%%%%%%%%%%%%%%%%%%%%%%%%%%%%%%%%%%%%%%%%%%%%%%%%%%

Our main conclusions may be summarized as follows:
\begin{itemize}
    \item Visual inspection of our deep HST/ACS imaging reveals no obvious stellar counterpart within our HST/ACS imaging. Using {\tt ArtPop} to generate resolved galaxies following the local stellar mass–size relation with simulated HST observing conditions, we can visually rule out a faint galaxy counterpart down to $M_{\star} \approx 10^{3.5} \ M_{\odot}$.
    \item Quantitatively, 10,000 CMD realizations of faint systems, mimicking the HST observations and applying identical selection cuts, indicate that a $10^{4} \ M_{\odot}$ stellar counterpart can be ruled out at 99.5\% confidence within a projected radius of 180~pc centered on Cloud-9. This radius matches the size of a Leo T-like ($M_{\star}\sim 10^{5} \ M_{\odot}$) galaxy at the distance of M94.
    \item A similar analysis using simulated CMDs yields an average upper stellar mass limit of $10^{3.5} \ M_{\odot}$, in excellent agreement with our visual comparison between the HST/ACS imaging field and simulated observations.
    \item Extending the search radius to account for a potential offset of up to $\sim$1~kpc between the VLA peak emission and any stellar counterpart results in no statistically significant detection above the background level, as determined by analyzing sources on the opposing (WFC2) chip. 
\end{itemize}

We suggest that future work on Cloud-9 should focus on three complementary fronts. First, deeper optical observations, particularly with JWST, could further lower the upper limit on its stellar mass. Second, numerical simulations should explore whether the slightly perturbed morphology of Cloud-9 is consistent with RELHICs experiencing ram-pressure stripping. Third, deep H$\alpha$ imaging could probe the expected ring-like emission in its outer regions~\citep{Sykes2019}. Together with existing radio data, these observations would enable joint constraints on the intensity of the local UVB––which may be influenced by Cloud-9’s proximity to M94–– while also providing an opportunity to map the dark matter distribution beyond Cloud-9's core.

n the $\Lambda$CDM framework, the existence of a critical halo mass scale for galaxy formation naturally predicts galaxies spanning orders of magnitude in stellar mass at roughly fixed halo mass. This threshold marks a sharp transition where galaxy formation becomes increasingly inefficient~\citep{Benitez-Llambay2020}, yielding outcomes that range from halos entirely devoid of stars to those able to form faint dwarfs, depending sensitively on their mass assembly histories. Even if Cloud-9 were to host an undetected, extremely faint stellar component, our HST observations, together with FAST and VLA data, remain fully consistent with these theoretical expectations. Cloud-9 thus appears to be the first known system that clearly signals this predicted transition, likely placing it among the rare RELHICs that inhabit the boundary between failed and successful galaxy formation. Regardless of its ultimate nature, Cloud-9 is unlike any dark, gas-rich source detected to date.

%Even if Cloud-9 were to host an undetected, extremely faint stellar component, our HST  observations, together with those from FAST and VLA, are fully consistent with theoretical expectations of a threshold mass scale, $M_{\rm crit}$, below which galaxy formation becomes increasingly inefficient~\citep{Benitez-Llambay2020}. In this framework, $M_{\rm crit}$ marks the sharp transition between dark and luminous halos, where outcomes range from halos that remain entirely devoid of stars to those able to form dwarf galaxies, depending sensitively on their mass assembly histories. Cloud-9 appears to be the first known system that clearly signals this predicted transition, likely placing it among the rare RELHICs that inhabit the boundary between failed and successful galaxy formation. Regardless of its ultimate nature, Cloud-9 is unlike any dark, gas-rich source detected to date.

\begin{acknowledgments}

We thank the anonymous referee, as well as Jacco van Loon and Yakov Faerman, for useful comments which improved the quality of our work. ABL acknowledges support by the Italian Ministry for Universities (MUR) program “Dipartimenti di Eccellenza 2023-2027” within the Centro Bicocca di Cosmologia Quantitativa (BiCoQ), and support by UNIMIB’s Fondo Di Ateneo Quota Competitiva (project 2024-ATEQC-0050). JFN acknowledges the hospitality of the Max-Planck Institute for Astrophysics, the Donostia International Physics Centre, and Durham University during the completion of this manuscript. Support for program GO-17712 was provided by NASA through a grant from the Space Telescope Science Institute, which is operated by the Association of Universities for Research in Astronomy, Inc., under NASA contract NAS5-26555.

\end{acknowledgments}

The HST/ACS data is available at this MAST DOI: \url{http://dx.doi.org/10.17909/10gr-vg66}.

\facilities{HST (ACS)}
\software{DOLPHOT \citep{Dolphin2000, Dolphin2016}, DrizzlePac \citep{2015ASPC..495..281A}}

\bibliography{paper}{}
\bibliographystyle{aasjournalv7}

\end{document}